%% file: paper.tex
\documentclass[sigconf]{acmart} 

\input{configs}

\begin{document}

\title{\systemName: Failure-Resilient Model Serving for Resource-Constrained Edge Environments}



\begin{abstract}
\input{sections/0-abstract}
\end{abstract}

\maketitle

\fancyhead[R]{Wu et al.}

\section{Introduction}
\label{sec:introduction}
\input{sections/1-introduction}

\section{Background and Motivation}
\label{sec:background}
\input{sections/2-background}

\section{\systemName Design}
\input{sections/3-design}





\section{Implementation}
\label{sec:implementation}
\input{sections/4-implementation}

\section{Evaluation}
\label{sec:evaluation}
\input{sections/7-evaluation}

\section{Discussion}
\label{sec:discusion}
\input{sections/10-discussion}

\section{Related Work}
\label{sec:related_work}
\input{sections/8-related_work}

\section{Conclusion}
\label{sec:conclusion}
\input{sections/9-conclusion}

\begin{acks}
We thank the anonymous SoCC reviewers for their valuable insight and feedback. This research is supported by National Science Foundation (NSF) grants 2213636, 2105494, 2211302, 2211888, 2325956, 23091241, the U.S. Department of Energy Award DE-EE0010143, US Army contract W911NF-17-2-0196, and support from VMware. This work used Amazon Web Services through the CloudBank, which is supported by NSF grant 19250001. 
\end{acks}

\bibliographystyle{ACM-Reference-Format}
\bibliography{paper}

\end{document}

%% file: configs.tex
\usepackage{tikz}
\usepackage{amsmath}

\usepackage{filecontents}

\usepackage{graphicx} 
\usepackage{subcaption}  
\usepackage{array}
\usepackage{multirow} 
\usepackage{multicol}
\usepackage{siunitx}
\usepackage{listings}
\usepackage{xspace}
\usepackage[]{hyperref}
\usepackage{circledsteps}
\usepackage{float}
\usepackage{enumitem}
\usepackage{mathtools}
\usepackage[linesnumbered, noend]{algorithm2e}

\def\systemName{\texttt{FailLite}\xspace}

\newcommand\lilly[1]{\textcolor{black}{#1}}
\newcommand\walid[1]{\textcolor{black}{#1}}

\author{Li Wu}
\affiliation{
  \institution{University of Massachusetts Amherst}
  \country{liwu@cs.umass.edu}
}
\author{Walid A. Hanafy}
\affiliation{
  \institution{University of Massachusetts Amherst}
  \country{whanafy@cs.umass.edu}
}

\author{Tarek Abdelzaher}
\affiliation{
  \institution{University of Illinois at Urbana-Champaign}
  \country{zaher@illinois.edu}
}

\author{David Irwin}
\affiliation{
  \institution{University of Massachusetts Amherst}
  \country{irwin@ecs.umass.edu}
}

\author{Jesse Milzman}
\affiliation{
  \institution{Army Research Laboratory}
  \country{jesse.m.milzman.civ@army.mil}
}

\author{Prashant Shenoy}
\affiliation{
  \institution{University of Massachusetts Amherst}
  \country{shenoy@cs.umass.edu}
}

\copyrightyear{2025}
\acmYear{2025}
\setcopyright{cc}
\setcctype{by}
\acmConference[SoCC '25]{ACM Symposium on Cloud Computing}{November
19--21, 2025}{Online, USA}
\acmBooktitle{ACM Symposium on Cloud Computing (SoCC '25), November
19--21, 2025, Online, USA}\acmDOI{10.1145/3772052.3772243}
\acmISBN{979-8-4007-2276-9/2025/11}

\ccsdesc[500]{Computer systems organization~Reliability}
\ccsdesc[500]{Computer systems organization~Redundancy}
\ccsdesc[500]{Computer systems organization~Distributed architectures}

\keywords{Edge Computing, Model Serving, Resilience AI Systems,  Fault-Tolerance}

%% file: sections/0-abstract.tex

Model serving systems have become popular for deploying deep learning models for various latency-sensitive inference tasks. While traditional replication-based methods have been used for failure-resilient model serving in the cloud, such methods are often infeasible in edge environments due to significant resource constraints that preclude full replication. To address this problem, this paper presents FailLite, a failure-resilient model serving system that employs (i) a heterogeneous replication where the failover model is a smaller variant of the original one, (ii) an intelligent approach that uses warm replicas to ensure quick failover for critical applications while using cold replicas, and (iii) progressive failover to provide low mean time to recovery (MTTR) for the remaining applications. We implement a full prototype of our system and demonstrate its efficacy on an experimental edge testbed and large-scale simulations. Our results using 27 models show that FailLite can recover all failed applications with 2$\times$ lower MTTR and only a 0.6\% reduction in accuracy. Under extreme failure scenarios, where 50\% of edge sites fail simultaneously, FailLite improves recovery rate by at least 39.3\% compared to the baseline methods.


%% file: sections/1-introduction.tex
In recent years, deep learning-based systems have revolutionized many domains, such as autonomous driving, augmented reality,  personal assistants, and Internet of Things (IoT) analytics~\cite{Li2020:EdgeAI, Satya2021:TheRole, Li2024:Acies-OS}. The widespread adoption of deep learning models is attributed to the growing availability of accelerators, such as NVIDIA GPUs and Google TPUs. To train a Deep Neural Network (DNN) model, users often rely on cloud data centers (e.g., Amazon AWS and Google Cloud) to access their powerful DNN accelerators via simple application interfaces. For example, services like Amazon SageMaker allow users to upload their data and use a simple user interface to select the models to train~\cite{AmazonSageMaker}. In contrast to DNN training, which is largely a batch workload, runtime use of trained models---often called model serving or model inference---tends to be latency-sensitive in nature.
\lilly{To meet these low-latency requirements of such applications, many users deploy models on edge data centers located closer to end devices~\cite{Satya09_Cloudlets, Satya17_emergence}. Recent advances in edge accelerators, such as edge GPUs and TPUs, have further enabled edge computing to support latency-sensitive models serving workloads effectively~\cite{Liang2023:Queueing, Satya2021:TheRole}. }

With the advances in AI, researchers have addressed many challenges in building model-serving systems. For instance, researchers have addressed the latency challenges by optimizing the network overheads~\cite{Hanafy2023:Understanding, Huang2020:CLIO}, processing latency 
\cite{Daniel2017:Clipper, Gujarati2020:Clockwork, Soifer2019:MSTInference, Zhang2019_HeteroEdge, Samplawski2020:Towards, Zhang2021:DeepSlicing, Liang2023:Queueing}, cost \cite{Zhang202:MArk, Ahmad2024:Loki, Liang2020:AIEge, Fan2019:OnCost}, energy efficiency~\cite{Liang2023:Delen, Hanafy2021:DNNSelection, Wan2020:ALERT, Zadeh2020:GOBO}, and addressing challenges of workload dynamics~\cite{Ahmad2024:Proteus, Ahmad2024:Loki, Zhang2020:ModelSwitching}.
\lilly{However, failure resiliency in model serving--particularly in resource-constrained environments like the edge--has been largely overlooked. Yet, edge environments are especially prone to hardware failures, resource contention, and connectivity issues, which can significantly disrupt model serving. As a result, building systems that can tolerate such failures is essential for resilient edge deployments.}

The traditional approach for ensuring fault tolerance of an application task is to replicate it on additional servers and to fail over to a backup server when the primary fails. This approach to crash fault tolerance is commonly used for critical cloud-based services.  Cloud-based model serving, which is useful when tight latency requirements are not needed, has also employed replication to satisfy demand and latency constraints and to ensure failure resiliency. 
For instance, a recent effort \cite{Soifer2019:MSTInference} utilized backup requests to multiple replicas and cross-server cancellation to ensure that all requests are processed within their latency requirements.
Such replication may be feasible in cloud settings where resources are abundant and can be scaled on demand, but edge environments are often highly resource-constrained, which may preclude replication of deployed models for fault tolerance. Moreover, edge computing clusters are often deployed across multiple geographically distributed locations and may face many single points of failure issues, where the entire cluster becomes out of service or unreachable. 

Designing a failure-resilient model serving system at the edge involves several key challenges. \lilly{First, such a system must operate under tight resource constraints, where traditional approaches like full model replication and failover are often impractical. Second, because model serving is typically interactive and latency-sensitive, the system must minimize mean time to recovery (MTTR) to reduce user-perceived disruptions during failures. Third, the system must tolerate more severe scenarios, such as multiple server failures or the loss of an entire edge site, by rapidly restoring services at other suitable edge locations.} 

\walid{To address this problem, we present \systemName, a failure-resilient model serving system for edge computing clusters. 
The core insight of \systemName is to provide fault tolerance using the concept of {\em heterogeneous replication}, where we use a smaller variant of the original model as a failover replica, ensuring high resiliency without incurring a high resource overhead of conventional replication techniques. By leveraging accuracy-resource trade-offs offered by deep learning model families~\cite{Hanafy2021:DNNSelection, resnet, Tan2020:EfficientNet} and generations~\cite{Redmon2016:YOLO, Reis2024:YOLOv8}, \systemName utilizes the DNN models' flexibility in increasing the resiliency of model serving applications by selecting failover backups according to the available resources. During resource-constrained periods, \systemName can mitigate failures by trading resiliency for accuracy loss. Conversely, when resources are sufficient, it can serve queries using more accurate models to improve accuracy. 
}

\lilly{Our second insight is that applications can vary in terms of their criticality and failure resilience needs, allowing  \systemName to intelligently choose the level of failure resilience on a per-model basis. Specifically, \systemName uses a two-step proactive and progressive failover approach. For highly critical models, \systemName uses proactive replication with warm replicas, which offer negligible recovery time. For less critical models, it employs cold replicas with a progressive failover strategy. In this case, \systemName first loads the smallest model variant to minimize the MTTR, then progressively transitions to a more accurate variant to improve the accuracy. This hybrid approach allows \systemName to maintain a low recovery time while still achieving high accuracy during failover, even under tight resource constraints.}

In designing, implementing, and evaluating \systemName, our paper makes the following contributions:

\begin{enumerate}[leftmargin=*]
    \item \lilly{We present the design of \systemName, a failure-resilient model serving system for resource-constrained edge environments. 
    \systemName uses a novel two-step failover approach that uses warm replicas for critical models and progressive failover with cold replicas for less critical ones. FailLite leverages heterogeneous replication by using smaller model variants as backups and progressively loading more accurate variants to minimize MTTR and improve accuracy.}
    \item We implemented \systemName as a framework-agnostic control plane with failure detection and failure-resilient policies. We integrated \systemName with the NVIDIA Triton Inference Server, an open-source model serving framework, to show the utility of our design. We also integrated our \systemName with a custom-built discrete event simulation platform for large-scale evaluations.
    \item Our experimental evaluation of \systemName of a real test bed using 27 deep learning models shows that in resource-constrained environments,  \systemName can recover all failed applications with 2$\times$ lower
    MTTR, only for a 0.6\% reduction in accuracy. 
    In addition, our large-scale simulations with 69 DNN models show that \lilly{under extreme failure scenarios, where 50\% of edge sites fail simultaneously, \systemName improves recovery rates by at least 39.3\% compared to the full-size model replication baselines.}

\end{enumerate}

%% file: sections/2-background.tex
In this section, we provide a background on model serving, failures, fault-tolerance techniques, and trade-offs in DNN models. Lastly, we motivate the need for \systemName by highlighting the key challenges in designing failure-resilient model serving systems.

\subsection{Model Serving on the Edge}
Edge Computing brings cloud-like computational and storage resources to the network's edge and provides users with low-latency applications~\cite{Satya17_emergence, Satya09_Cloudlets}. Although edge computing was pioneered more than a decade ago, the rise of AI made it more critical due to the resource requirements and performance objective of AI applications that often go beyond the most sophisticated on-device capabilities~\cite{Satya2021:TheRole, Li2020:EdgeAI}. 
The model serving process, also known as model inference, can be described as follows: a user or a sensor node sends input data $x_i$ to a specific ML inference application, and the application responds with output data $y_i$, where $y_i = f(x_i)$. The function $f(\cdot)$ may represent a single model inference or a pipeline of different ML models. Users often rely on model serving frameworks (e.g., Nvidia Triton~\cite{nvidia_triton} and Kubeflow~\cite{kubeflow2025}) to run their DNN models as they provide many management and monitoring capabilities.
Model serving services benefit from edge servers' low latency, the availability of powerful DNN accelerators, and data privacy capabilities to provide users with real-time processing capabilities. 
For example, applications such as IoT analytics, video streaming, and AR/VR systems heavily rely on edge resources to cover their high computing demand~\cite{Nexus, Salehe2019:VideoPipe}.

\subsection{Network and Resource Failures}
Network and compute resources are prone to multiple types of failures, such as transient, crash, or byzantine failures, impacting the model serving systems' reliability~\cite{Hanafy2023:FailureRes, koren2007faultbook}. Resilient execution ensures service availability and correctness in the event of faults and aims to minimize metrics such as downtime or mean time to recovery (MTTR). 
\walid{In this work, we focus on network and server crashes and transient failures that make an edge server or a cluster permanently or temporarily inaccessible or unreachable.
Crash failures result from various factors, including human errors, power outages, battery drain, hardware malfunctions, software misconfigurations, and security attacks~\cite{superbench2024ATC}.
Transient failures in networks and servers could occur from benign reasons, such as network congestion or CPU and IO bottlenecks, and malicious, such as jamming and Denial of Service Attacks.
Addressing crash and transient failures is critical in edge computing since they are irrecoverable and more likely to occur than in cloud computing, as edge computing resources typically operate in a less controlled environment with limited to no resource redundancy.} 

\walid{Researchers have discussed only failures in distributed training \cite{He2023:DNNTrainingFail, Li2017:Understandingerror}. However, in the context of model serving systems, little or no work has addressed the resiliency challenges and the effect of server or network crashes. To our knowledge, researchers have only highlighted the benefits of special DNN architectures (e.g., multi-exit and skip-connection) for failure resiliency~\cite{yousefpour2019guardians, yousefpour2020resilinet, majeed2022continuer} without addressing the design and mechanisms of applying these techniques for failure resiliency systems. 
Nonetheless, we analyze their behavior in \autoref{sec:evaluation}.} 
Finally, we note that addressing Byzantine failures from adversarial clients (e.g., DNN Perturbation attacks \cite{Lin2020:Threats}) is outside the scope of this work. 

\begin{figure}
    \centering
\includegraphics[width=\linewidth]{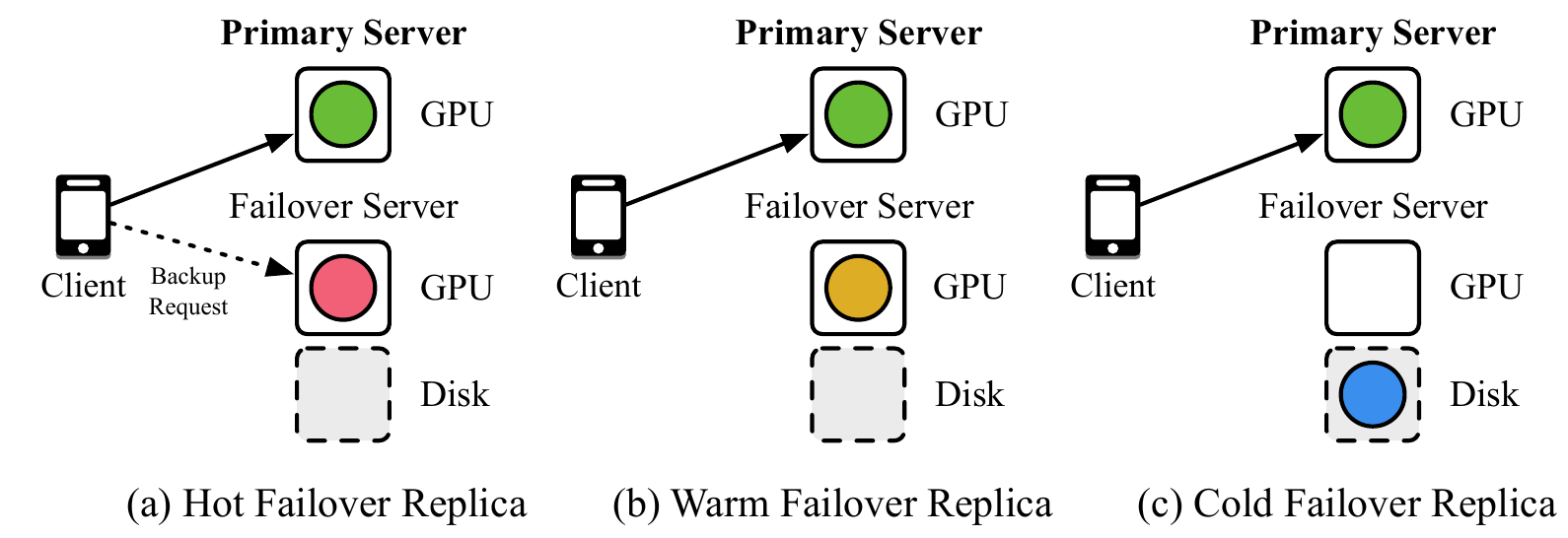}
    \caption{Approaches to utilize a failover replica.}
    \label{fig:bg_failover_replica}
\vspace{-7mm}
\end{figure}

\subsection{Failover Replication}
Replication is a key strategy in ensuring resilient systems, where components are backed with a failover replica. In the case of model serving systems, researchers have used traditional failover replica approaches, where a model can have a hot (also referred to as active-active), warm (active-passive), or cold failover backup. \autoref{fig:bg_failover_replica} highlights the difference between hot, warm, and cold backups. For instance, in \cite{Soifer2019:MSTInference}, the authors have used a hot backup (see \autoref{fig:bg_failover_replica}a) where each request is issued twice, and the extra request is canceled once a response is available. To avoid wasting compute cycles, warm backups are used where the DNN models are loaded to memory but do not process any requests.
In contrast to hot and warm backups, which are loaded to memory, cold backups are just cached to disk and loaded to memory only when the failure occurs.

Despite the advantages and trade-offs involved with traditional failover replication techniques, edge environments are resource-constrained, and such failover replication methods fall short in resilient execution. For example, using a hot backup needlessly increases the system's load and average response time, while resources may not allow all applications to have a warm backup. Moreover, unlike cloud resources, adding more resources to edge environments introduces high-cost overheads and may not be feasible due to space and power constraints. To overcome such resource limitations, one approach is to limit the failure recovery to a set of critical applications or reduce the quality of service, commonly known as {\em graceful service degradation} \cite{graceful_2023, uni-processor-graceful, Hanafy2023:FailureRes, defcon}. Our proposed \systemName utilizes this method to increase the resiliency of model serving systems by inter-playing the accuracy-resiliency trade-offs. 

\subsection{DNN Models Trade-offs}
\begin{figure}[t]
  \centering%
    \begin{subfigure}[t]{0.7\linewidth}%
    \includegraphics[width=\textwidth]{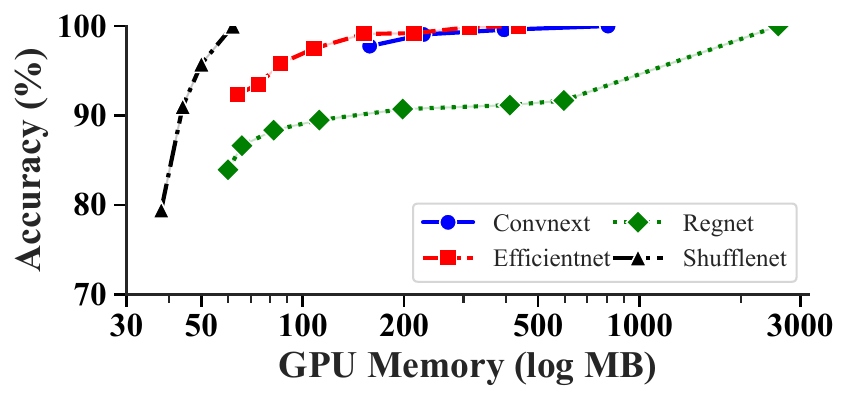}%
    \caption{Resource-Accuracy Trade-offs}%
    \label{fig:models_tradeoffs_acc}%
  \end{subfigure}%
  \\
  \begin{subfigure}[t]{0.7\linewidth}%
    \centering%
    \includegraphics[width=\textwidth]{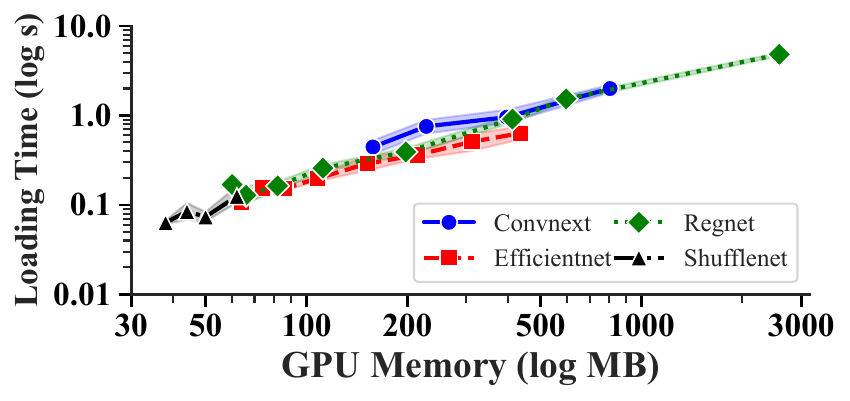}%
    \caption{Loading Time}%
    \label{fig:models_tradeoffs_loading}%
  \end{subfigure}%
    \caption{Accuracy-Resource Tradeoff (a) and Loading time (b) across DNN models.} 
    \label{fig:models_tradeoffs}
    \vspace{-5mm}
\end{figure}

Recent research has highlighted the accuracy-memory trade-offs of DNNs~\cite{Hanafy2021:DNNSelection, Ahmad2024:Proteus, Zhang2020:ModelSwitching, Halpern2019:OneSize, Liang2023:Delen, Ahmad2024:Loki, Li2023:Clover}, where increases in accuracy come at an exponential increase in resource requirements.
~\autoref{fig:models_tradeoffs} demonstrates the accuracy-memory trade-offs and the loading time across different pre-trained model families from Pytorch~\cite{pytorch_models}. 
~\autoref{fig:models_tradeoffs_acc} shows the trade-off between model size and accuracy (we report normalized accuracy to the biggest model) in four different model families. As shown, the figure highlights that, \lilly{in many cases,} significant reductions in memory requirements introduce limited reductions in accuracy. For example, ConvNext-Tiny is 5.1$\times$ smaller compared to  ConvNext-Large but only reduces accuracy by 1.89\%. ~\autoref{fig:models_tradeoffs_loading} also highlights a key aspect of DNNs where loading time is a function of memory size, where accurate models are typically large, thus requiring higher loading time. 



This unique trade-off has motivated users to address runtime issues.
For example, the authors of \cite{Ahmad2024:Proteus, Zhang2020:ModelSwitching} have used model switching to address workload dynamics by replacing the DNN with a smaller variant at peak hours, allowing all requests to abide by their required SLO. Moreover, Clover \cite{Li2023:Clover} used model switching to reduce the energy consumption when the energy is generated from brown energy sources. In contrast to these approaches that focus on workload dynamics, \systemName utilizes model variants in optimizing the resiliency of model serving systems by using a compressed or smaller size failover replica, which significantly decreases the resource requirements for the failover replicas, an approach we denote as {\em heterogeneous replication}.

\subsection{Failure Resilience Challenges on the Edge} \label{sec:bg_motivation}
\systemName leverages the DNN model's flexibility to ensure resilient model serving in case of crash failures in resource constrained edge environments. Designing \systemName involves addressing the following research challenges:

\begin{enumerate}[leftmargin=*]
\item[\texttt{\bf C1}] \textbf{Model Selection.} The first challenge is model selection; naively choosing the smallest variant unnecessarily reduces accuracy. In contrast, selecting larger or identical replicas restricts the capacity to provide failover replicas. Therefore, the system must weigh the trade-offs across applications and other factors when choosing the failover models.
    
\item[\texttt{\bf C2}]\textbf{ Balancing Accuracy-MTTR Trade-offs.} The second challenge is that while offering failover backups for all applications highly decreases the MTTR, adding backups for all applications in resource-constrained scenarios unnecessarily reduces the accuracy for applications that fail since resources are allocated to backups of functioning applications. In contrast, cold backups enable us to load the most accurate models, but activating cold backup models takes more time, significantly increasing the MTTR, as loading large models from disk is a lengthy process (see ~\autoref{fig:models_tradeoffs_loading}).

\item[\texttt{\bf C3}]\textbf{ Model Placement.} The third challenge is that the system must optimize the placement decisions while considering a unique set of constraints while optimizing the selection decisions. The system must adhere to the performance requirements of both the primary and failover models, as well as the resource limits. Additionally, the placement must address correlated failures, where co-located edge servers are more likely to fail in a correlated manner.

\item[\texttt{\bf C4}]\textbf{Failure Detection and Fast Failover.} Finally, the system must promptly detect failures and implement the recovery plan, quickly restoring applications’ functional behavior. Furthermore, the system must notify clients of the new location and the selected DNN model.
\end{enumerate}

%% file: sections/3-design.tex
To address the above challenges, we propose \systemName, a failure-resilient model serving system for resource-constrained edge environments. \walid{\systemName employs} a two-step proactive and progressive failover approach, which intelligently provides failover replicas in a proactive and dynamic manner to minimize the MTTR while ensuring minimal accuracy degradation. In this section, we first give an overview of the two-step approach, then detail each of the two steps, and lastly explain how these two steps sit together within \systemName and how to extend them to tolerate geographically correlated edge failures.


\subsection{\systemName Overview}

\begin{figure}[t]
    \centering
    \includegraphics[width=\linewidth]{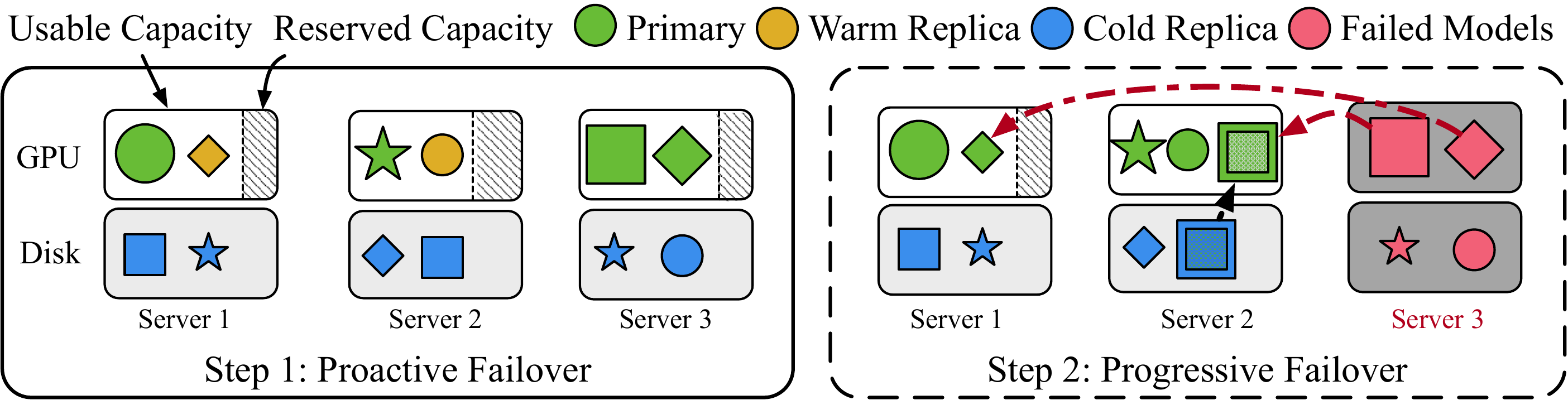}
    \caption{Overview of \systemName's two-step approach. \lilly{We note that the same shape represents the same applications, and size reflects the model variant}.}
    \label{fig:system_overview}
    \vspace{-7mm}
\end{figure}

The key insight of \systemName is leveraging the flexibility in DNN models by using the concept of \emph{heterogeneous replication}, where we allow the failover backup to be a smaller model, reducing the failover replica resource requirements. This heterogeneous replication allows \systemName to increase the number of failover replicas, enhancing the end-to-end model serving resiliency. However, as highlighted in \autoref{sec:bg_motivation}, designing a fault tolerance model serving system requires addressing issues of model selection and placement and balancing the accuracy-MTTR trade-offs. To do so,   \systemName features a two-step fault tolerance process, illustrated in ~\autoref{fig:system_overview}. As shown, initially, \systemName places warm failover replicas for a set of critical applications while considering available resources and performance requirements. 
In the second step, \systemName uses a progressive failover process in which cold failover replicas are loaded in an iterative manner from disk to memory based on the servers that have failed.
As we will demonstrate in \autoref{sec:evaluation}, our proposed two-step approach overcomes the limitations of singular methods (i.e., only using warm or cold backups). Our hypothesis is that \emph{using our two-step approach enables \systemName to maximize accuracy during edge server failures by preloading warm backups for critical applications while reserving some capacity for cold backups. Additionally, the progressive loading of cold backups on demand allows \systemName to enhance the system's accuracy while minimizing MTTR.} 
Next, we provide an overview of these two steps. 

\subsubsection*{\textbf{1) Proactive Failover}}
\systemName uses a configurable parameter to split the free capacity between proactive and the progressive backups (see ~\autoref{fig:system_overview}). \lilly{By taking into account both the usable capacity (free capacity - reserved capacity) and applications that require a warm backup (e.g., critical applications), \systemName selects model variants for those critical applications while allowing all applications to have a cold backup (i.e., backups stored on servers’ disks). In particular, \systemName optimizes the warm backup model selection and placement to maximize the total accuracy while ensuring the successful placement of the warm backups.  These warm backups are preloaded into memory and can serve model inference requests instantly, minimizing downtime when primary servers fail. }




\subsubsection*{\textbf{2) Progressive Failover}} 
When a server fails, \systemName needs to determine and load models for the application without a warm backup in a dynamic fashion. 
The key challenge in this step is that the model selection and placement becomes a real-time decision and highly affects the MTTR. In addition, loading cold backups often has a high loading and warm-up time, a function of the model size, as illustrated in \autoref{fig:models_tradeoffs_loading}. To overcome this challenge, we propose a progressive model selection and placement approach.
\systemName first uses a greedy heuristic that maximizes the accuracy under available capacity to perform backup model selection and placement. Next, \systemName loads these models in a progressive manner. It instantaneously loads the smallest model variant per application to minimize the MTTR and then loads the selected larger model to maximize the accuracy. 
Although the application may experience higher performance degradation when using the smallest model, it greatly decreases the service downtime. As shown in \autoref{fig:models_tradeoffs_loading}, two model variants from the same model family can have up to an order of magnitude difference in their loading time. 

Taking the four applications in ~\autoref{fig:system_overview} as an example, \systemName loads warm backups for two applications while the others have cold backups. By reserving capacity for cold backups, \systemName can utilize server 2 for failing over applications from servers 1 and 3. If server 3 fails, \systemName switches the two applications to a warm backup on server 1 and to a cold backup on server 2, respectively.

\subsection{Proactive Failover}\label{sec:design_proactive}
Our proactive fault tolerance step employs users' preference (e.g., using application criticality level) as well as the required fault tolerance degree of non-critical applications.
\systemName assumes that users define a set of applications that must have a warm failover replica while allowing all models to have cold failover replicas. However, naively allocating all available capacity to warm backups may curb the overall resiliency of the systems, as \systemName may utilize oversized failover replicas, leaving no room for other applications. To address this issue, \systemName reserves some of the capacity for cold replicas determined by a configuration parameter.

\noindent\textbf{Warm Backup Model Selection and Placement Problem.} We formulate our warm backup model selection and placement as an Integer Linear Programming (ILP) optimization problem. 
Our optimization problem considers $N$ applications running on $S$ edge servers and a set of applications $K: K\subseteq N$ that must have a warm backup. We assume that the primary instances are placed and each server has a free capacity $c_k^r$ where $r$ represents different resource types (e.g., GPU memory, CPU RAM, and compute cycles). 
We assume that users or system administrators define a system-wide failover parameter $\alpha$, where a higher $\alpha$ reserves more resources for cold backups, improving the overall system's resilience, while a smaller $\alpha$ reserves more resources for critical applications, maximizing their accuracy.
Furthermore, our settings assume that each application $i$ has a set of model variants $n_i$ with varying resource demands $d_{ij}^r$ and accuracy $a_{ij}$. We note that since different applications have different accuracy ranges, we normalize the accuracy, where $a_{ij} = a_{ij}/max(a_{ij})$, a common approach in such situations~\cite{Ahmad2024:Proteus}. The latency of running variant $j$ of application $i$ on server $s_k$ is denoted as $l_{ijk}$.  
The notations used in the problem are summarized in ~\autoref{tab:notations}.
We define our problem as a maximization problem as follows:

{
    \begin{equation}
    \label{eq:objective} \max_{\substack{x_{ijk}^*}} \quad
    \sum_{i \in K}   \sum_{j\in n_i} \sum_{k\in S}  a_{ij} \cdot q_{i} \cdot  x_{ijk}  
    \end{equation}
}

\noindent s.t.
{
\begin{align}
\label{eq:resource}
    & \sum_{i \in K} \sum_{j\in n_i} x_{ijk} \cdot d_{ij}^r \leq c_k^r, \quad \forall k, \ \forall r &\\
    \label{eq:reserved_resource}
    & \sum_{i \in K} \sum_{j\in n_i} \sum_{k\in S} x_{ijk} \cdot d_{ij}^r \leq (1-\alpha) \sum_{k\in S} c_k^r, \quad \ \forall r &\\
    \label{eq:independence}
    & \sum_{j \in n_j} x_{ijp_i} = 0,  \quad \forall i &\\
    \label{eq:replica}
    & \sum_{j\in n_j} \sum_{k \in S} x_{ijk} = 1, \quad \forall i &\\
    \label{eq:latency}
    &\sum_{j \in n_j} \sum_{k \in S} l_{ijk} \cdot x_{ijk} \leq L_i,  \quad \forall i &\\
    \label{eq:binary}
    &x \in \{0, 1\}&
\end{align}
}

The objective function (\autoref{eq:objective}) maximizes the effective accuracy ~\cite{Ahmad2024:Proteus, Zhang2020:ModelSwitching} by considering the normalized accuracy and \walid{maximum} applications' request rate $q_i$ by only considering the set of applications $K$.
\autoref{eq:resource} ensures the total resource demands of all warm failover backups assigned to a server do not exceed its available capacity. 
\autoref{eq:reserved_resource} ensures that the capacity reserved for cold backups is respected.
\autoref{eq:independence} is a primary backup independence constraint, where failover replicas do not share the same server as the primary. 
\autoref{eq:replica} ensures each application has one backup. \autoref{eq:latency} limits backup placement to servers and variants that meet the latency SLO. Lastly, ~\autoref{eq:binary} ensure that $x_{ijk}$ is binary.

\begin{table}[t]
\caption{Notations used in \systemName.}
\label{tab:notations}
\resizebox{\linewidth}{!}
{%
\begin{tabular}{c|l}
\toprule
\textbf{Notation} & \textbf{Definition}  \\ \toprule
$N$ & set of applications.\\
$S$ & set of servers.\\
$K: K \subseteq N$ & set of applications that require a warm backup.\\ 
$\alpha \in [0, 1]$ &  fault-tolerance parameter.\\ 
$c_{k}^r$ & resource capacity of server $k$ of type $r$.\\ 
$n_i$ & set of models for application $i$.\\
$a_{ij}$ & normalized accuracy of model $j$ of application $i$.\\ 
$d_{ij}^r$ & resource demand of model $j$ of application $i$ of type $r$.\\ 
$p_i$ & primary server of application $i$.\\  
$q_i$ & \walid{maximum} request rate of application $i$.\\  
$l_{ijk}$ & latency of running model $j$ of application $i$ on server $k$.\\  
$L_i$ & latency constraints of application $i$.\\\midrule
$x_{ijk} \in \{0, 1\}$ & 1 when model $j$ for application $i$ assigned to server $k$.\\ \bottomrule
\multicolumn{2}{l}{where, $i \in K$, $j\in n_i$, and $k \in S$.}\\
\end{tabular}%
}

\vspace{-3mm}
\end{table}

\subsection{Progressive Failover}

In the proactive model selection and placement step, \systemName leverages an ILP to select and place model variants across servers. 
Despite the similarities between the proactive approach 
in \autoref{sec:design_proactive}, which can be used to select and place the failover backups by considering the subset of affected applications $K'$, the location of cold replicas, and the available resources. The key challenge in formulating this problem as an ILP is that solving ILPs typically takes large amounts of time. 
Although this may be suitable for the proactive step, it highly increases the MTTR, a key objective of \systemName. Moreover, as mentioned earlier, cold backups often exhibit a high loading and warm-up time, which can be in orders of seconds (see \autoref{fig:models_tradeoffs_loading}).

\input{sections/heuristic_alg}

To address these challenges, we propose a model selection and placement heuristic, described in \autoref{alg:algorithm}, and a progressive model loading approach. When \systemName detects a failure, \autoref{alg:algorithm} takes the list of impacted applications that have no warm backup to fail over and available servers to select the model variants (Line 2-6) and to decide their placement (Line 7-14). First, it computes the available resource capacity and maximum resource demands and computes a resource ratio $\delta$, where $\delta\geq1$ denotes that resources are mostly sufficient to place full backups for all applications, while $\delta<1$ means that some applications may face accuracy degradation. \autoref{alg:algorithm} then selects the model variant that matches the $\delta$ (e.g., when $\delta$ is 0.5, the match( ) function selects a model whose resource demands are 50\% of the full-size model). The algorithm then iterates over the applications and their model variants (from the selected model to smaller) and tries to place them \walid{across servers in a memory worst-fit manner, where applications are placed on servers with the highest available memory capacity, until a feasible server and model variant is found.}
Lastly, \autoref{alg:algorithm} relies on our worst-fit placement and uses the available memory to upgrade the selected model if the selected server can fit a more accurate model. After determining the model variant and its placement, \systemName first loads the smallest variant, ensuring a low MTTR, and after the connection is restored, it upgrades it to the selected model.

\subsection{Putting It Together}\label{sec:design_together}

~\autoref{fig:architecture} presents the architecture of \systemName, featuring a centralized controller that implements failure detection and our proposed two-step failover policies. \systemName \lilly{uses an agent sitting in an edge server} to facilitate the two-step failover process.  Here, we present the workflow of \systemName and demonstrate how the proactive and progressive failover approaches are \lilly{combined}. 

First,  when a new application arrives (\Circled[inner color=black]{1}), \systemName uses the predetermined placement of primary replica (e.g., by the lowest latency or user selection) and uses the proactive failover approach to select and place warm backup models.
In either case, \systemName informs the \systemName agent to retrieve the model from the disk and load it to the GPU memory according to the policy (\Circled[inner color=black]{2}).
At runtime, \systemName uses a heartbeat signal to determine the state of the model serving servers. 
When \systemName detects a failure, it follows the progressive failover approach and determines the location of the cold backup using our proposed heuristic approach(\Circled[inner color=black]{3}). In the case of cold backups, \systemName progressively loads the model, allowing the system to have both a low MTTR and a high accuracy(\Circled[inner color=black]{4}). Once the failover process completes, \systemName informs the load balancer or clients of the new location to reroute their inference requests(\Circled[inner color=black]{5}). \walid{Lastly, once the system functionality is restored, \systemName redirects inference requests back to the primary models.}

\subsubsection*{\textbf{Extending \systemName to Geographically Correlated Failures.}}
Although the aforementioned policy focuses on single failure tolerance, geographically corrected failures within or across edge sites are commonplace in edge environments. To tolerate these failures, \systemName's proactive failover can allow applications to have multiple warm backups (i.e., updating \autoref{eq:replica} as a multi-replica constraint). Moreover, \systemName can extend the primary independence constraint (\autoref{eq:independence}) to edge site independence, where primary and failover backup do not share the same edge cluster or create an exclusion list where applications in an edge site are only allowed to have backups in a subset of the listed edge sites. Importantly, our progressive failover approach can adapt to these widespread failures by selecting models and placing them on the fly. In \autoref{sec:eval_failures}, we evaluate \systemName behavior in edge site failures.

%% file: sections/heuristic_alg.tex
\RestyleAlgo{ruled}
\begin{algorithm}[t]
    \caption{\texttt{FailLite\_Heuristic()}}
    \label{alg:algorithm}
    \KwIn{Affected Applications $N'$, Available Servers $S'$.}
    \KwOut{Model Selection $X$ and Placement $Y$}
    \textbf{Initialization:} $X \gets \{\}$ and $Y \gets \{\}$\;%
    $C^r = \sum_{k}^{S'} c_k^r$ \tcp{Compute Available Capacity.}
    $D^r = \sum_{k}^{N'} d_{max}^r$ \tcp{Compute Max Demand.}
    $\delta = C^r/D^r$ \tcp{Compute demand ratio.}
    \For {$i\in N'$}{%
        $X[i] = $ match($n_i,\delta$) \tcp{Select variants close to $\delta$.}
    }
    \For {$i\in N'$}{
    \tcp{Iterate over model variants}
    \For {$j\in [X[i]..1]$}{
        $k = $WorstFit($j, S'$)\tcp{Model Placement}
        \If{$k \neq \phi$}{%
        \tcp{Feasible placement was found.}
            $Y[i] = k$\;
            $X[i] = j$
        }
    }
    }
   \For {$i\in N'$}{%
        \tcp{Increase model accuracy.}
    $X[i] = $upgrade\_model($X[i], Y[i]$)\;
   }
    \Return $X, Y$
\end{algorithm}

%% file: sections/4-implementation.tex
\begin{figure}[t]
    \centering
    \includegraphics[width=1.0\linewidth]{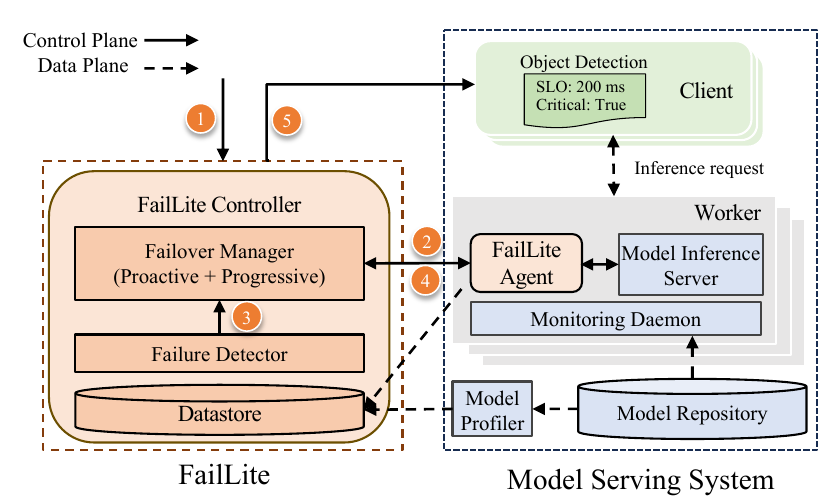}
    \caption{Overview of \systemName Architecture.} 
    \label{fig:architecture}
    \vspace{-7mm}
\end{figure}

In this section, we detail the prototype implementation of \systemName. 
Our current prototype controller is integrated with the Nvidia Triton Inference Server~\cite{nvidia_triton}. However, it is not limited to Triton's design or interfaces. We implemented \systemName using Python in $\sim$5k SLOC. \systemName is open source and is available at https://github.com/umassos/faillite.
In the remainder of this section, we will detail the implementation of \systemName and how the proposed architecture is integrated with Triton.

\subsubsection*{\bf \systemName Failure Detection} 
At the heart of \systemName, is our failure detection approach. 
The \systemName agent utilizes a periodical \emph{push alive message}, sent every $T$ms, which our experiments use $T=20 \ ms$. If the controller does not receive two consecutive messages, it initiates our proposed failover process. 

\subsubsection*{\bf \systemName Failover} We implemented our proposed two-step failover approach, where it places warm backups for critical applications (i.e., proactive failover) and reacts to failure detection by loading the failover replicas for other applications. We implemented our proposed ILP for the proactive placement step using the Python interface of Gurobi v12.0.0.
In contrast, our progressive failover heuristic \autoref{alg:algorithm}) is implemented in Python and initiated to load models for applications without a warm backup. Lastly, \systemName asynchronously informs the applications' client of the failover inference server and model using push notifications implemented using websockets v15.0.

\subsubsection*{\bf \systemName Data Store.}
\systemName utilizes a data store to collect system state (e.g., utilization), performance metrics (e.g.,  response time), and application profiles (e.g., memory requirements and service time) across model variants. \systemName also maintains the locations of the primary and backups of applications. Our implementation utilizes Redis v7.4.2, an in-memory key-value data store, which allows replication and periodic checkpoints, enhancing the resiliency of our controller.

\subsubsection*{\bf Integrating \systemName and Triton.}
We build the worker node on top of Nvidia's open-source enterprise model inference system, NVIDIA Triton Inference Server v24.12-py3~\cite{nvidia_triton}, and TensorRT v10.8. We run Triton as a Docker container using Docker v27.5.1, where we allocate all accelerator resources to the container. Triton can serve multiple models within one server, collect system and inference metrics, and support model management at runtime. Triton utilizes a local model repository where models are placed on the disk before loading. To load/unload models, Triton presents a \texttt{Load/Unload\{ID\}} API, where ID is the application ID, and each application model variant has a unique ID. To differentiate the applications, each inference model is indexed with appID and model variant (e.g., $AppID\_MVar$). 

Our implementation augments the Triton server with two key features. First, with failure detection capabilities, where \systemName agent implements a periodic heartbeat, where it pushes a keep-alive message every \lilly{20 ms}. Second, \walid{\systemName employs a local agent that dynamically configures, using Triton \texttt{Load/Unload\{ID\}} API, the running models to enhance the system's resiliency.} 
\lilly{To do so, \systemName's controller uses the agent to load multiple variants of an inference model on a Triton inference server, based on the failover policy. In addition, \systemName uses the agent to perform progressive failover. Specifically, during the progressive failover, \systemName first instructs Triton to load the smallest model and redirects requests to it to minimize downtime. Meanwhile, it uses Triton API to load the more accurate model in the background. Once the more accurate model is ready, FailLite switches over to it and lets Triton unload the smaller one.} Lastly, we note that since Triton assumes that all models are locally available, our \systemName agent also manages the local model repository by fetching the needed model variants (e.g., from the cloud).\footnote{We assume that servers typically have ample storage and may maintain many cold replicas. Although addressing storage limitations is beyond the scope of this paper, one solution is to assume that servers rely on a cloud-based model repository where they download models on the fly or use a replication where application models are replicated on multiple servers or sites.}

%% file: sections/7-evaluation.tex
In this section, we evaluate the performance of \systemName and our proposed policies using a real-world testbed. We also augment our results with large-scale simulations highlighting the supremacy and the scalability of our solution. In doing so, we answer the following research questions:

\begin{itemize}[leftmargin=*]
    \item How does \systemName react to an edge server failure? How does our two-step approach address the limitation of traditional failover approaches? 
    \item What is the performance of \systemName with different resource constraints, application configurations, model families, and edge site failures. 
    
    \item What is the overhead and scalability of \systemName?
\end{itemize}

\subsection{Experiment Setup}
\subsubsection*{\textbf{Experimental Testbed}} 
We deploy \systemName on a local testbed consists of 7 Dell PowerEdge R630, each equipped with a 40-core Xeon E5-2660v3 CPU running on Ubuntu 22.04. Each server has 256 GB of memory, a 400 GB Intel 730 SSD, and up to 10 Gbps networking speed. Additionally, 6 of these servers include an NVIDIA A2 GPU with 1280 CUDA cores and 16 GB of GPU memory. We assume these six servers represent three edge sites, each with two servers, and we use a server as a controller node.

\subsubsection*{\textbf{DNN Workloads}}
We evaluate \systemName with 16 model families PyTorch  ~\cite{pytorch_models}, totaling 69 DNN models. PyTorch models include their accuracy and model requirements (number of parameters and FLOPs) across these models and different architectures. \walid{While our approach is not limited to specific model grouping and scales,} we assume that each architecture represents a family, and we calculate the accuracy by normalizing the accuracy to the most accurate model within a family. 
We categorize these model families into three classes: small, medium, and large, according to their maximum resource demands, the difference between the largest and smallest models. 
Our experiments first focus on deploying a number of DNN models from 5 model families, including Mobilenet, Shufflenet, Convnext, Efficientnet, and Regnet. We later expand our analysis to include all model families using simulation.  

We compile these models using TensorRT and use Triton's Model Analyzer to profile all the model variants. Triton Model Analyzer measures the GPU memory, compute utilization, and average inference time for batch size 1. 
Lastly, we note that our approach can utilize static analysis techniques to estimate the memory and processing times~\cite{paleo, justus2018, cai2017neuralpower}. 

\subsubsection*{\textbf{\systemName Deployment}}
We deploy \systemName on the edge testbed. Each server acts as a worker node and serves inference requests. Each server hosts a Triton inference server within a container and uses TensorRT as a backend. Furthermore, we disable the auto completion and set the model loading thread to 10 to reduce the model loading overhead introduced by Triton. Each worker node reports the heartbeat to the controller every 20 ms. The \systemName controller runs on one edge server and detects edge failures every 100 ms. 

In our real experiments, we first deploy the primary of these DNN applications using the worst-fit algorithm, resulting in around 50\% utilization in GPU memory and GPU. Our setup deploys a client on the same server as the primary. 
Moreover, unless otherwise mentioned, we assume that users provide a set $K$ of critical applications that must have a warm failover replica, which we set as 50\%, and we set the fault tolerance parameter $\alpha$ as 0.1.
Lastly, to make results comparable across settings, we fix the applications and control the available capacity via controlling a headroom parameter, which we range from 50\% to 10\%, representing different resource constraints, where unless otherwise mentioned, we assume that headroom is 20\%. 
Moreover, we extend the evaluation of \systemName through large-scale simulations. We utilize all the model families from PyTorch~\cite{pytorch_models} and expand the setup to 100 servers. We use the model profiles and MTTR numbers collected from the real testbed to ensure the fidelity of our results. To overcome the scalability issues of the ILP solver, our solutions utilize the proposed heuristic (\autoref{alg:algorithm}).

\subsubsection*{\textbf{Failure Injection}}
We inject failures of edge servers by stopping the Triton inference container. When the container is down, the \systemName Agent stops sending the heartbeat to the controller. 
In our experiments, we run model inference in the normal state for 60 seconds before injecting failures and collect another 60 seconds of data afterward.

\subsubsection*{\textbf{Evaluation Metrics}} We focus on three metrics to evaluate the performance of \systemName: (i) \textit{Recovery Rate (\%)}: The percentage of DNN applications that were affected by the failure and were recovered, (ii)\textit{ Mean Time To Recovery (MTTR) (s)}: The duration between the failure detection and notifying the client of the new application location, which includes model loading time for cold backups. Note that we only consider applications that were recovered; non-recovered applications have an infinite MTTR, and (iii) \textit{Accuracy Reduction (\%)}: The accuracy reduction of backup compared to primary. Note that we only consider applications that were recovered; non-recovered applications have zero accuracy.

\subsubsection*{\textbf{Baselines}} To the best of our knowledge, no existing model serving systems address the failure resiliency in resource-constrained environments. Therefore, we compare \systemName to the traditional approaches in failover replication that do not consider the accuracy-resource trade-off and try to use full-size models. Note that since these policies do not consider different model variants, they have no accuracy loss, but not all applications can be recovered. In doing so, we consider four baselines:

\begin{itemize}[leftmargin=*]
    \item \textbf{Full-Size-Warm}: This policy places full-size warm backups to minimize the MTTR of DNN applications. It first considers the set of critical applications $K$, then tries to place a full-size backup for the others. 
    \item \textbf{Full-Size-Cold}: This policy considers all applications and places cold backups. In case of failure, this policy first loads backups for the $K$ critical applications, then randomly loads other failover replicas. 
    \item \textbf{Full-Size-Warm ($K$)}: This policy considers both warm and cold backups, where it only places warm backups for the $K$ critical applications, while allowing all applications to have cold backups.
    \item \textbf{Split-Model}: \walid{Lastly, in ~\autoref{sec:eval_early_exit}, we compare to a policy inspired by the design in \cite{majeed2022continuer} and adopts a split and early-exit processing strategy, distributing DNN models across multiple servers. This architecture enhances resiliency by enabling the early stages of inference to operate independently and produce partial (less accurate) results, even when downstream components become inaccessible.}
    
\end{itemize}


\begin{figure}[t]
    \centering
    \includegraphics[width=0.9\linewidth]{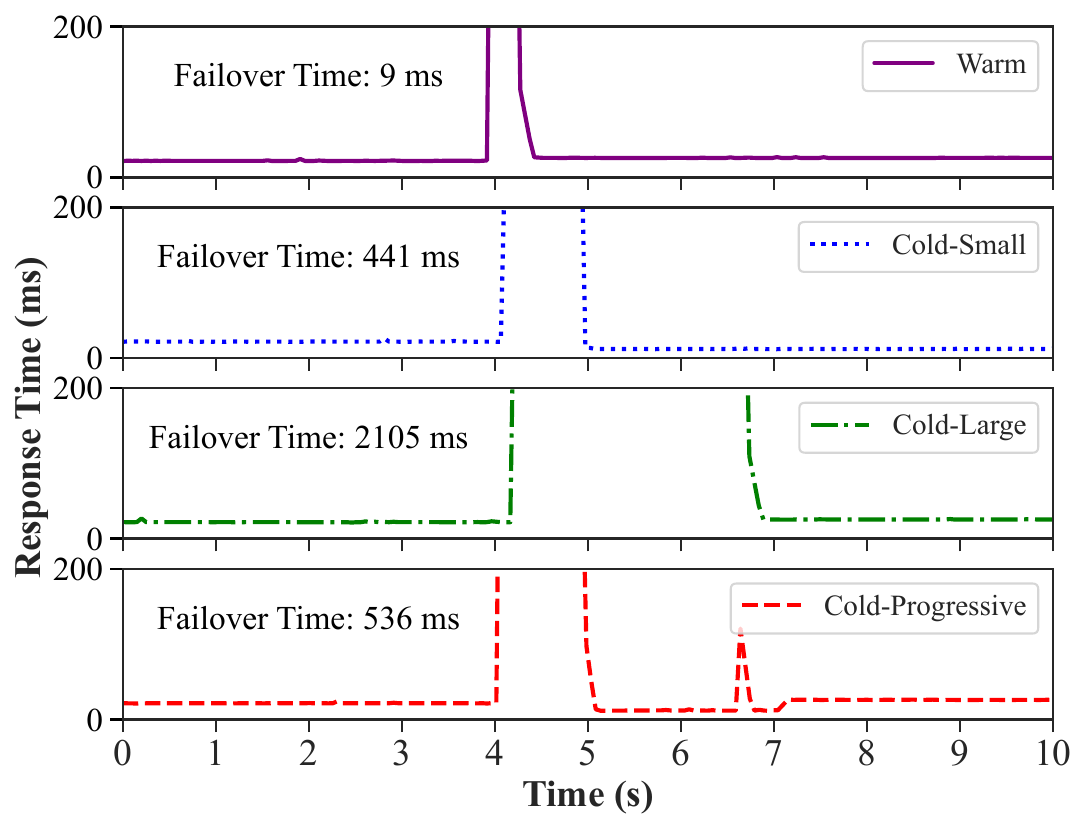}
    \caption{\systemName behavior across different types of backups, showing the advantages of warm backups and our proposed progressive failover.}
    \label{fig:behavior_single}
    \vspace{-7mm}
\end{figure}

\begin{figure*}[t]
  \begin{subfigure}[t]{0.32\textwidth}%
    \centering%
    \includegraphics[width=\textwidth]{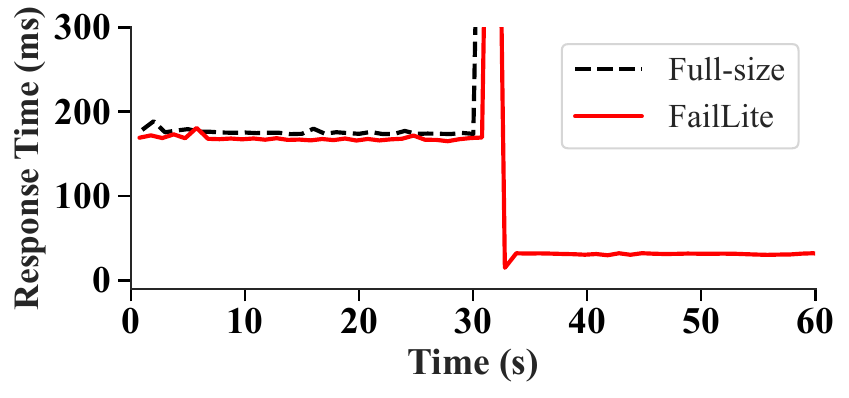}%
    \caption{Warm backup (small)}%
    \label{fig:behavior_multiple_warm}%
  \end{subfigure}%
  \begin{subfigure}[t]{0.32\textwidth}%
    \centering%
    \includegraphics[width=\textwidth]{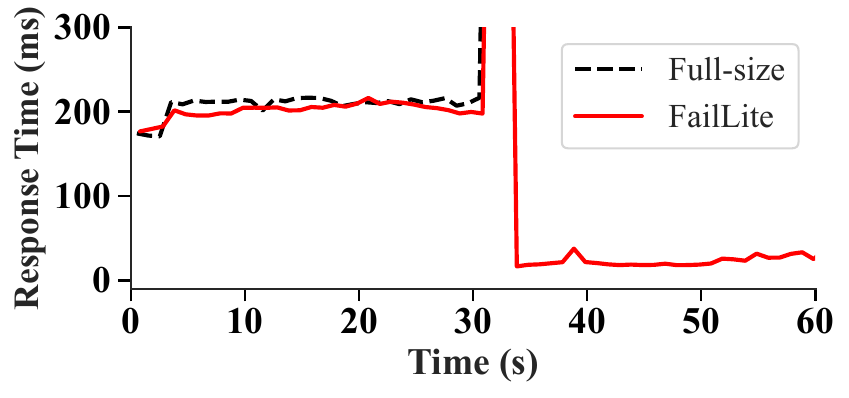}%
    \caption{Cold backup (small)}%
    \label{fig:behavior_multiple_cold_small}%
  \end{subfigure}%
  \begin{subfigure}[t]{0.32\textwidth}%
    \centering%
    \includegraphics[width=\textwidth]{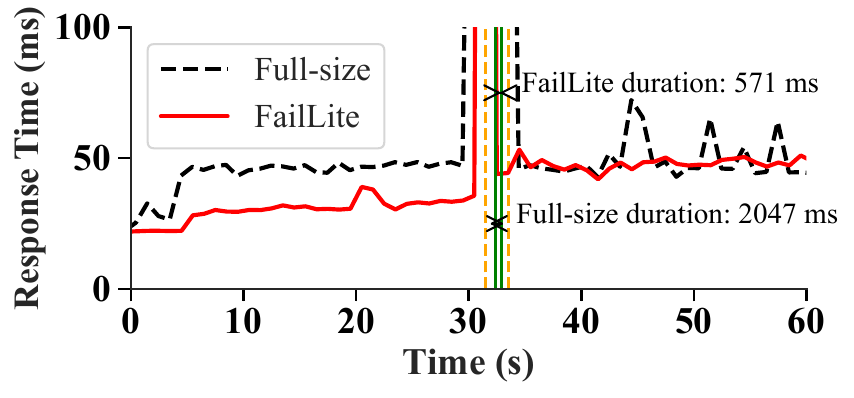}%
    \caption{Cold backup (full-size)}%
    \label{fig:behavior_multiple_cold_large}%
  \end{subfigure}%
  \caption{Comparing \systemName to Full-Size-Warm($K$), showing scenarios, where only \systemName can have a backup (a) and (b) as well as benefits of progressive failover in reducing the recovery time when using the full-size cold backup.}
  \label{fig:behavior_multiple}
  \vspace{-3mm}
\end{figure*}

\begin{figure}[t]
  \centering%
    \begin{subfigure}[t]{0.23\textwidth}%
    \includegraphics[width=\textwidth]{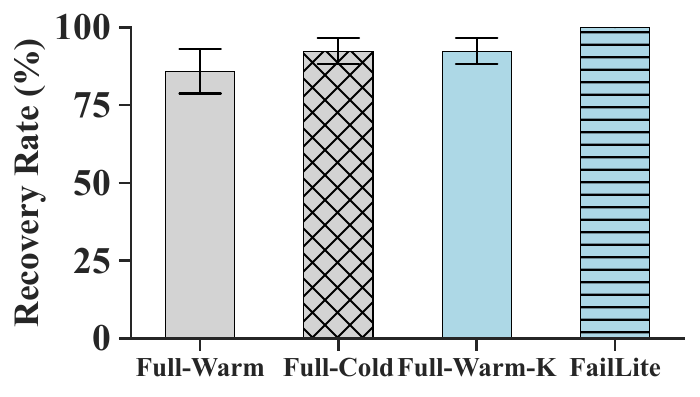}%
    \caption{Recovery Rate (\%)}%
    \label{fig:behaviour_summary_rate}%
  \end{subfigure}%
  \hfill%
  \begin{subfigure}[t]{0.23\textwidth}%
    \centering%
    \includegraphics[width=\textwidth]{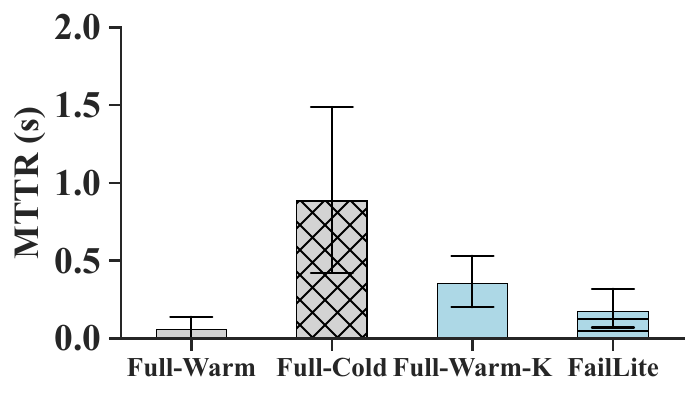}%
    \caption{MTTR (s)}%
    \label{fig:behaviour_summary_MTTR}%
  \end{subfigure}%
  \caption{Recovery rate and MTTR of \systemName. The accuracy reduction is 0.6\%.} 
  \label{fig:behaviour_summary}
  \vspace{-7mm}
\end{figure}

\subsection{\systemName in Action}
First, we evaluate the behavior of \systemName in failure recovery, where we consider a simple scenario where a single application is running and show the failure recovery process across different approaches.  ~\autoref{fig:behavior_single} shows the behavior of \systemName when consider a model serving application using the Convnext model family. The x-axis represents the experiment type, where we inject the failure at the 4th second, and the y-axis shows the clients' response time. The figure shows the difference in behavior across types of failover backups, where a warm backup (aside from its size) will have a small failover time, where clients can quickly restore their behavior and reissue the queued requests after failure, restoring the normal behavior after $\sim$\lilly{10} ms. In contrast, when using the cold backups, the model loading time (see ~\autoref{fig:models_tradeoffs_loading}) is dependent on the model size. For example, ~\autoref{fig:behavior_single} shows that the small model of size 158MB, requires \lilly{441} ms to load, while the large one of size 806MB, requires \lilly{2105} ms to load, increasing the MTTR between small and large by a 3.86$\times$. Lastly, the figure shows the benefits of our progressive approach, where it reduces the MTTR while retaining the accuracy of the large. We note that, in this progressive approach, the client is oblivious to the switch, where \systemName does the change using the same network interface, yielding a small spike in the response time.

Next, we extend our experimental scenario, where we consider 5 model families and load models to cover 50\% of the compute and memory resources, resulting in 46 different applications. \autoref{fig:behavior_multiple} shows the behavior of \systemName compared to the Full-Size-Warm($K$), which considers warm backups for $K=50\%$ critical application and assume that the headroom (usable space to load backups) is 20\%. \autoref{fig:behavior_multiple} shows the behavior of \systemName and Full-Size-Warm($K$) for three selected scenarios, highlighting the advantages of \systemName, but we summarize the results across all applications in ~\autoref{fig:behaviour_summary}. 
\autoref{fig:behavior_multiple_warm} shows a scenario where ~\systemName utilizes a warm backup of small size to circumvent the resource limitation, maintaining a small MTTR (85 ms), in a scenario where a full-sized model cannot be loaded. \autoref{fig:behavior_multiple_cold_small} shows another decision made by \systemName, where it decides only to load a smaller-sized model, allowing the application to be recovered despite the resource limitation. In this case, \systemName will face a 1316 ms MTTR, as loading the application from disk takes more time. Lastly,  \autoref{fig:behavior_multiple_cold_large} \lilly{illustrates a scenario where both FailLite and Full-Size-Warm (K) fall back to cold backup using full-size models. Despite using the same model, FailLite reduces MTTR by 72\% through progressive failover.}

\autoref{fig:behaviour_summary} shows the aggregate behavior of \systemName and three baseline policies in terms of recovery rate (\autoref{fig:behaviour_summary_rate}) and (\autoref{fig:behaviour_summary_MTTR}). We note that we do not report the accuracy reduction, as baselines use full-size models, leading to no accuracy reduction, while \systemName exhibits a 0.6\% accuracy reduction.
The figure reports the average recovery rate and MTTR across six runs where we fix the application placement and, each time, we inject a failure on one of our six servers. 

As shown in \autoref{fig:behaviour_summary_rate}, in all scenarios, \systemName was able to maintain a high recovery rate, where it was able to load a failover backup for all applications. 
In contrast, other baselines could not host a failover for all applications. \lilly{Alternative methods cannot provide backups for all applications under failure, resulting in variable recovery rates (e.g., 86\% coverage for Full-Size Warm, 94\% for Full-Size Cold)}. \autoref{fig:behaviour_summary_MTTR} completes the picture by showing the MTTR across baselines.
The figures highlight the trade-offs of the choices made by the Full-size baselines. 
For instance, the Full-Size (Warm) policy was able to provide only a failover replica (in case of failure) for 85.9\% and 76\% of the applications on average and in the worst case. In contrast, although the Full-Size (Cold) was able to overcome this limitation, \autoref{fig:behaviour_summary_MTTR} shows that this approach has a high average MTTR, 15.8$\times$ larger than that of the Full-size (Warm). Moreover, the figure shows that despite how Full-Size-Warm($K$) may address the limitation of either approach, it still falls short in comparison to \systemName. For instance, \autoref{fig:behaviour_summary_rate} shows how \systemName has a recovery rate higher by 7.7\% and an average MTTR better by  2$\times$.

\subsubsection*{\textbf{Key Takeaways}} 
\textit{\systemName proposed a two-step approach, heterogeneous replication, and progressive loading address the limitation of traditional failover techniques. 
Our results show that compared to Full-Size-Warm($K$), \systemName is able to maintain a 100\% recovery rate and decrease the MTTR by 2$\times$, only for a 0.6\% accuracy loss.}

\begin{figure*}[t]
    \centering
    \includegraphics[width=0.6\linewidth]{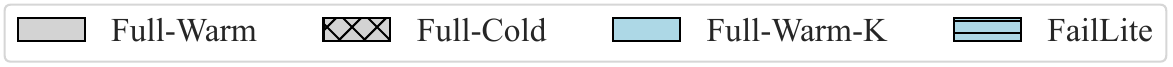}\\
    \hfill
  \centering%
    \begin{subfigure}[t]{0.3\textwidth}%
    \includegraphics[width=\textwidth]{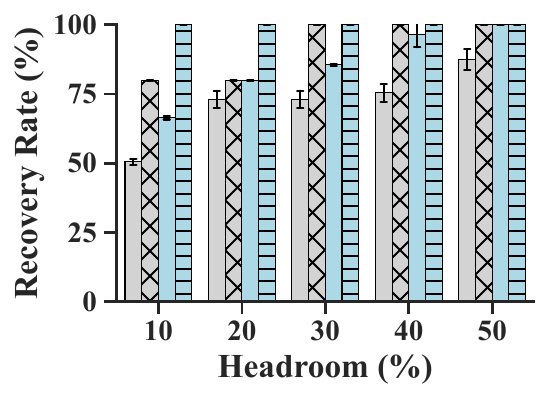}%
    \caption{Recovery Rate (\%)}%
    \label{fig:headroom_recovery_rate}%
  \end{subfigure}%
  \hfill%
  \begin{subfigure}[t]{0.3\textwidth}%
    \centering%
    \includegraphics[width=\textwidth]{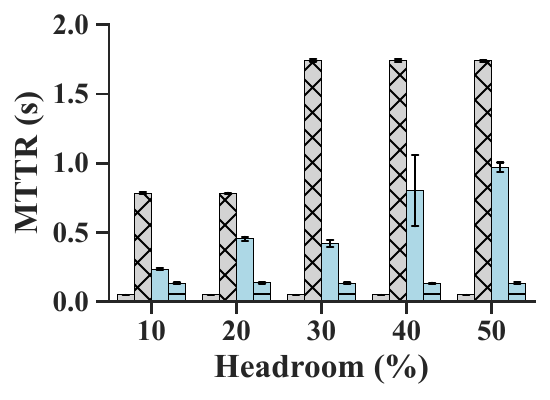}%
    \caption{MTTR (s)}%
    \label{fig:headroom_mttr}%
  \end{subfigure}%
  \hfill%
  \begin{subfigure}[t]{0.3\textwidth}%
    \centering%
    \includegraphics[width=\textwidth]{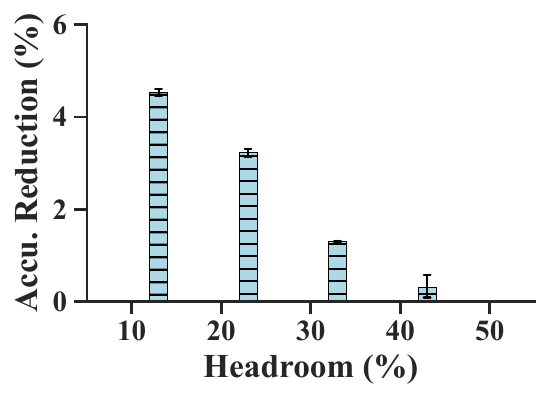}%
    \caption{Accuracy Reduction (\%)}%
    \label{fig:headroom_accuracy}%
  \end{subfigure}%
  \hfill
  \hfill
  \caption{Impact of resource constraints on the decisions and performance of \systemName and our baselines. We introduce resource constraints by changing the headroom available for failover backups.}
  \label{fig:headroom}
  \vspace{-5mm}
\end{figure*}

\subsection{Impact of Resource Constraints}
Despite the ability of \systemName to enhance failure resiliency in resource-constrained environments, the quantity of available resources influences the decisions made by \systemName. For example, when resources are already overloaded, there is no room to enhance the system's resiliency. In contrast, in low utilization scenarios, traditional failover techniques shall behave quite well. In this section, we evaluate the performance of \systemName and depict its ability to optimize its decisions in different resource contention scenarios. We simulate 10 edge sites with 100 servers in total and deploy the same mixture of DNN models as the real experiments across these servers. To make results comparable across settings, we fix the applications to 640 and control the available capacity via controlling a headroom parameter, which we range from 50\% to 10\%, representing different resource constraints. 


\autoref{fig:headroom} shows the behavior of \systemName and traditional failover baselines across different headroom settings. \autoref{fig:headroom_recovery_rate} shows the recovery rate across different settings, highlighting the ability of \systemName to recover all applications, even in highly constrained settings. In contrast, the traditional failover approach has struggled, especially for small headrooms. For instance, when headroom is 10\%, the full-warm, full-cold, and full-warm-k are only able to recover 50.5\%, 79.8\%, and 66\% of the applications. Nonetheless, increased headroom increases the recovery rate, where at 50\%, all approaches, except for Full-Warm, are able to recover 100\% of the applications.

In addition, \autoref{fig:headroom_mttr} shows the runtime behavior of different policies. As expected, the Full-Warm backup has a low and stable MTTR, as the headroom size has no effect when only considering warm backups, yielding a 50 ms average failover time. In contrast, policies that depend on cold backups (i.e., Full-Cold and Full-Warm-K) have a much higher increase in MTTR, as higher headroom allows for failover replicas for larger models, increasing the MTTR. For instance, the average MMTR increases from 784 ms to 1742 ms between 10\% and 30\% headroom. The figure also highlights how \systemName addresses the limitation in these policies, where \systemName achieves a low MTTR by allowing all applications within $K$ to have a warm backup while progressively loading cold failover models, resulting in $\sim$134 MTTR. 
Finally, ~\autoref{fig:headroom_accuracy} shows the accuracy; as noted earlier, utilizing full backups have no accuracy loss, and we only considered restored applications. As shown, the high recovery rate and smaller MTTR, come at a cost of accuracy reductions, where at a headroom of 10\% \systemName loses 4.52\% of the accuracy of the models.

\subsubsection*{\textbf{Key Takeaways}} \textit{Although resource limitation highly affects the failure resiliency of traditional failover approaches, \systemName was able to circumvent the key limitations of these approaches and achieve a 100\% recovery rate for a 4.52\% reduction in accuracy.}

\begin{figure}[t]
    \centering
    \includegraphics[width=0.8\linewidth]{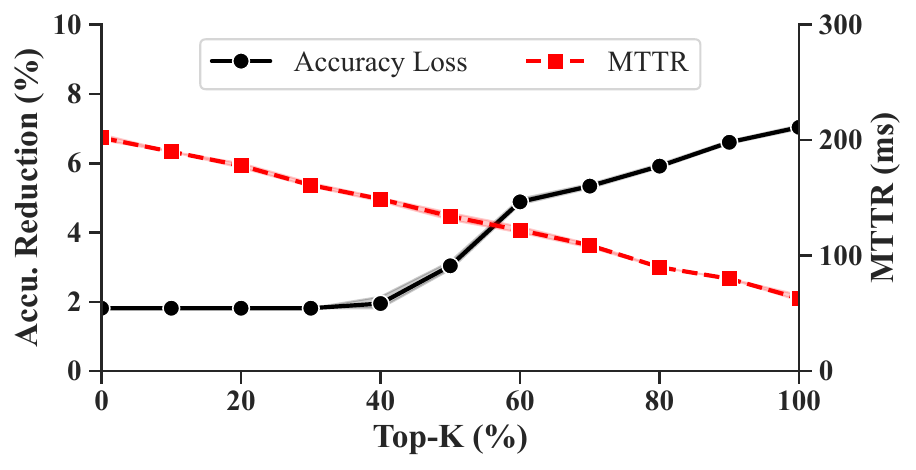}
    \caption{Impact of $K$.} 
    \label{fig:k_effect}
    \vspace{-7mm}
\end{figure}

\subsection{Impact of Applications' Criticality}
Another key factor in the decision of \systemName is the application criticality. \autoref{fig:k_effect} shows the effect of \systemName across different $K$ configurations, where we change $K$ from 0\% to 100\% using the large-scale simulation setup described earlier. As shown, the curve highlights the accuracy-MTTR trade-off in failure-resilient model-serving systems.
For instance, when all applications are critical ($K=100$), \systemName needs to place warm backups for all applications, reducing the overall accuracy. On the other hand, when none of the applications require a warm backup ($K=0$), \systemName can maximize the accuracy by loading the largest model for the affected applications at the cost of higher MTTR. Lastly, we note that although how users value different parameters is beyond the scope of this paper, we highlight that a balance point exists when $K=60\%$. 

\subsubsection*{\textbf{Key Takeaways}}
\textit{The decisions of \systemName introduces an accuracy-MTTR trade-offs, where the decisions of \systemName at $K$ introduces a low MTTR, while higher $K$, comes with a performance desegregation.}

\begin{figure*}[t]
   \centering%
      \includegraphics[width=0.6\linewidth]{figures/3_headroom_legend.pdf}\\
    \centering%
    \begin{subfigure}[t]{0.3\textwidth}%
    \includegraphics[width=\textwidth]{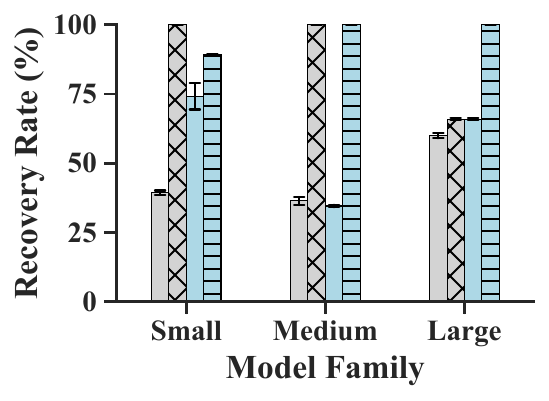}%
    \caption{Recovery Rate (\%)}%
    \label{fig:model_family_recovery}%
  \end{subfigure}%
  \hfill%
  \begin{subfigure}[t]{0.3\textwidth}%
    \centering%
    \includegraphics[width=\textwidth]{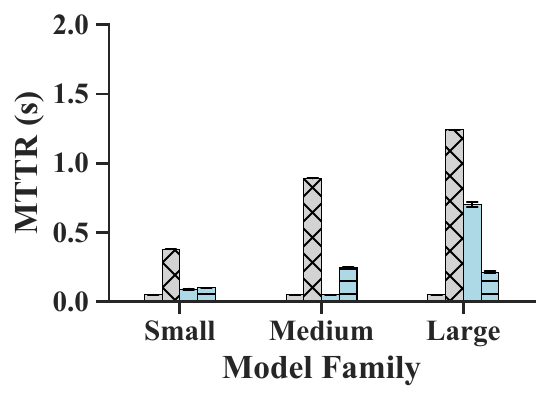}%
    \caption{MTTR (s)}%
    \label{fig:model_family_mttr}%
  \end{subfigure}%
  \hfill%
  \begin{subfigure}[t]{0.3\textwidth}%
    \centering%
    \includegraphics[width=\textwidth]{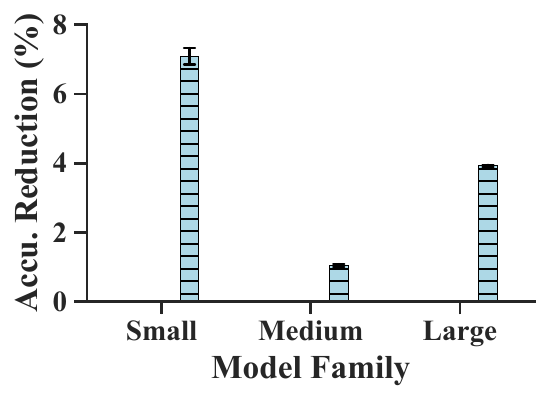}%
    \caption{Accu. Reduction (\%)}%
    \label{fig:model_family_acc}%
  \end{subfigure}%
  \caption{Impact of model family on the Recovery Rate (a), MTTR (b), and Accuracy Reduction (c).}
  \label{fig:model_family}
  \vspace{-5mm}
\end{figure*}


\subsection{Impact of Model Family}
We analyze the impact of different model families on the performance of \systemName. We categorize model families into three classes according to the maximal difference of resource demand between model variants: Small, Medium, and Large. For example, we consider Mobilenet model family as Small as the maximal resource demand difference is 12MB, and consider the Convnext model family as Large as its 648MB demand difference.
\autoref{fig:model_family} shows the performance of \systemName and other baselines when all applications are composed exclusively from one model family class. Lastly, we note that since model families have different resource requirements, the total number of applications changes across scenarios ranging between 3264 and 402 across the Small and Large classes.

\autoref{fig:model_family_recovery} shows that \systemName is able to achieve a higher recovery rate when the demand difference is bigger, while full-size baselines struggle. For instance, Full-Cold can only recover 65\% of the applications in the large class while \systemName can recover all. When the demand difference is small, \systemName can still outperform the Full-Warm-K policy by 15\% in recovery rate. \autoref{fig:model_family_mttr} also highlights the effect of model family class on the MTTR, where increases in model size introduce higher MTTR across all baselines. However, it is worth mentioning that even in the Large class, \systemName MTTR value was 214 ms on average.
\autoref{fig:model_family_acc} shows the accuracy reduction, although accuracy reduction is not comparable across scenarios, \systemName exhibits a maximum of 7.1\% reduction in accuracy. Lastly, as noted earlier, traditional policies have no accuracy loss. 

\subsubsection*{\textbf{Key Takeaways}}
\textit{With a larger difference in resource demand between model variants, \systemName is able to further optimize the resiliency of DNN models while achieving a 214ms MTTR.}

\begin{figure}[t]
   \centering%
      \includegraphics[width=0.8\linewidth]{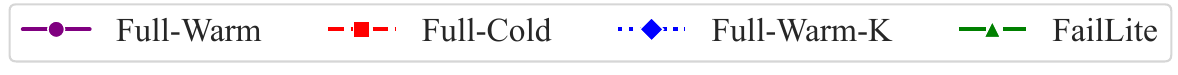}\\
  \centering%
    \begin{subfigure}[t]{0.23\textwidth}%
    \includegraphics[width=\textwidth]{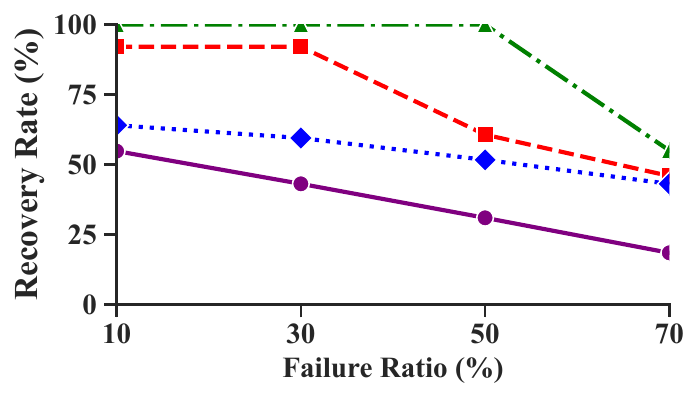}%
    \caption{Recovery Rate (\%)}%
    \label{fig:failure_ratio_rate}%
  \end{subfigure}%
  \begin{subfigure}[t]{0.23\textwidth}%
    \centering%
    \includegraphics[width=\textwidth]{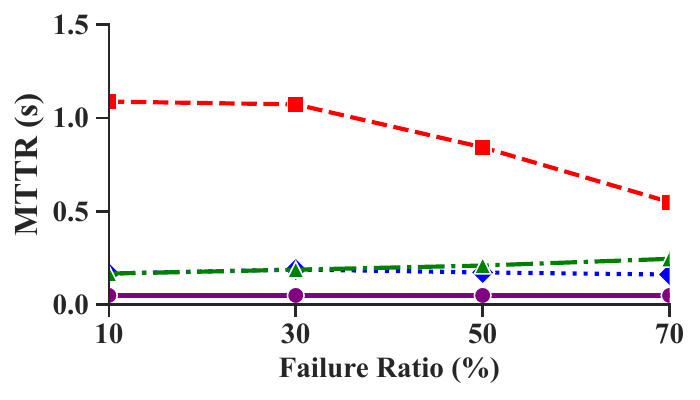}%
    \caption{MTTR (s)}%
    \label{fig:failure_ratio_MTTR}%
  \end{subfigure}%
  \caption{Impact of edge site failures.}
  \label{fig:failure_ratio}
  \vspace{-8mm}
\end{figure}

\subsection{Impact of Edge Site Failure}\label{sec:eval_failures}
Site-wide failures are common in the edge, as network and power systems do not have high levels of redundancy. To evaluate the effect of edge-wide failures, we utilize our large-scale simulation testbed of 100 servers and group these servers into ten sites of size 10. We then explore the effect of site failures by failing 1 - 7 of the 10 edge sites. Similar to previous experiments, we repeat these experiments multiple times to ensure the stability of our results. 
\autoref{fig:failure_ratio} shows the recovery rate and MTTR of \systemName and other approaches. We note that as mentioned in \autoref{sec:design_together}, we add a site independence constraint, where the warm backup is placed in a different site. As shown in \autoref{fig:failure_ratio}, \systemName is able to maintain 100\% failure recovery until 50\% of the sites fail enhancing the recovery rate by 7.9\% and 39.3\% compared to Full-Cold, for the single edge site and seven edge sites failures scenarios, respectively. \autoref{fig:failure_ratio_MTTR} also highlights how the MTTR changes across scenarios, where increases in the number of failed sites decreases the ability to load large models, decreasing the MTTR for all approaches. 

\subsubsection*{\textbf{Key Takeaways}}
\textit{The capabilities of \systemName can be extended to edge-site resilience. For instance, when 50\% of the edge sites fail, \systemName increases the recovery rate by at least 39.3\% compared to the full-size policies.}

\subsection{Comparing with Split-Model Approach}\label{sec:eval_early_exit}

\lilly{In this section, we compare \systemName with the the split processing approach, based on \cite{majeed2022continuer} and split processing across two servers. In our experiments, we utilize the ResNet architecture and add a split point and an exit layer that allows partial outputs. For comparison, \systemName uses the entire model as the full-size model, while using the first split with the exit layer as the smaller model variant.} 

\begin{figure*}[t]
  \centering%
  \hspace{6cm}\includegraphics[width=0.25\linewidth]{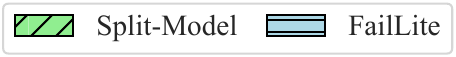}\\
\begin{subfigure}[t]{0.3\textwidth}%
    \centering%
    \includegraphics[width=\textwidth]{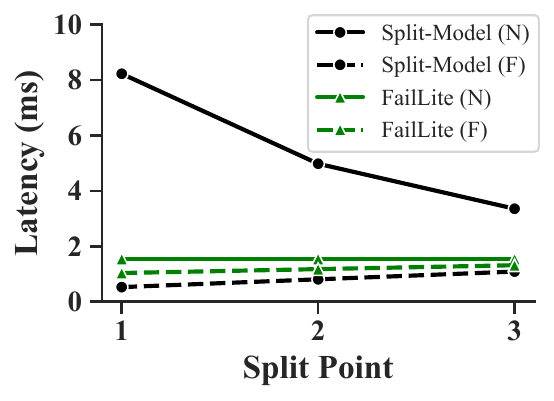}%
    \caption{End-to-end Latency (ms)}%
    \label{fig:multi_exit_latency}%
  \end{subfigure}%
  \hfill%
  \begin{subfigure}[t]{0.3\textwidth}%
    \centering%
    \includegraphics[width=\textwidth]{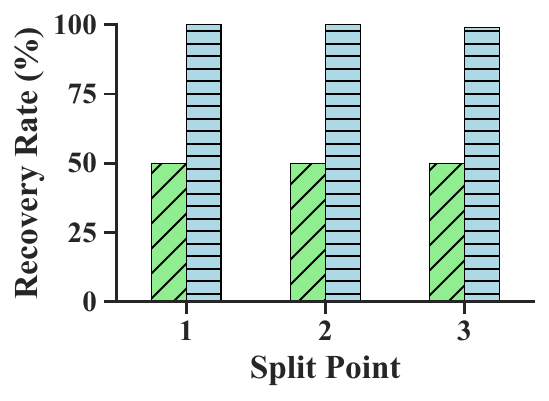}%
    \caption{Recovery Rate(\%)}%
    \label{fig:multi_exit_recovery_rate}%
  \end{subfigure}%
  \hfill%
  \begin{subfigure}[t]{0.3\textwidth}%
    \centering%
    \includegraphics[width=\textwidth]{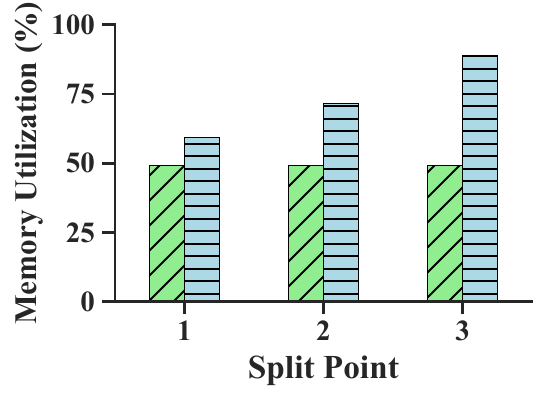}%
    \caption{Memory Utilization (\%)}%
    \label{fig:multi_exit_memory_usage}%
  \end{subfigure}%
  \caption{The performance of \systemName in comparison with split processing approaches, highlighting the latency (a), recovery rate (b), and memory utilization (c) of different approaches.
  }
  \label{fig:multi-exit}
  \vspace{-3mm}
\end{figure*}

\lilly{~\autoref{fig:multi-exit} demonstrates the impact of different split points on the behavior of \systemName and split processing approaches. 
~\autoref{fig:multi_exit_latency} shows the end-to-end latency of the two approaches under normal (N) and failure (F) scenarios. In normal cases, the split processing approach experiences a 5.3$\times$ higher latency compared to \systemName, as the split processing approach distributes exit models across servers, which introduces network and intermediate output transmission delay. 
~\autoref{fig:multi_exit_recovery_rate} shows \systemName achieve 100\% recovery rate while the split processing approach can only recover 50\% of the applications. This is because the split processing approach can only recover the application whose first exit model is alive, and \systemName can recover all applications by failing over to the smaller backup models.  
Lastly, \autoref{fig:multi_exit_memory_usage} illustrates the memory requirements of both approaches, where \systemName exhibits a higher memory overhead compared to the split processing approach, which utilizes warm failover backups and consequently consumes more memory. In particular, the figure shows that \systemName consumes 1.2$\times$ memory of the split processing approach at exit point 1, while consuming 1.8$\times$ at exit point 3, which provides a more accurate backup model.} 


\subsubsection*{\textbf{Key Takeaways}}
\textit{\systemName achieves a higher recovery rate and lower latency compared to the split processing approaches.}


\begin{figure}[t]

  \centering%
    \begin{subfigure}[t]{0.15\textwidth}%
    \includegraphics[width=\textwidth]{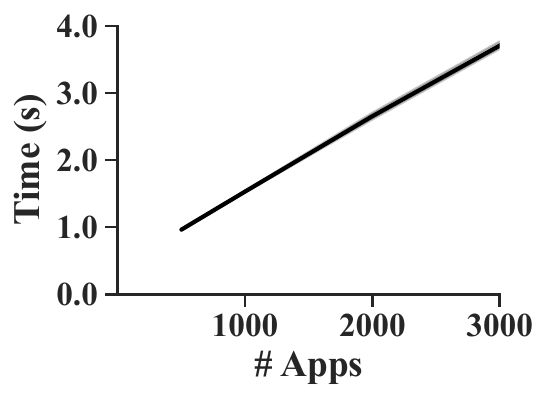}%
    \caption{\# Apps}%
    \label{fig:scalability_apps}%
  \end{subfigure}%
  \hfill%
  \begin{subfigure}[t]{0.15\textwidth}%
    \centering%
    \includegraphics[width=\textwidth]{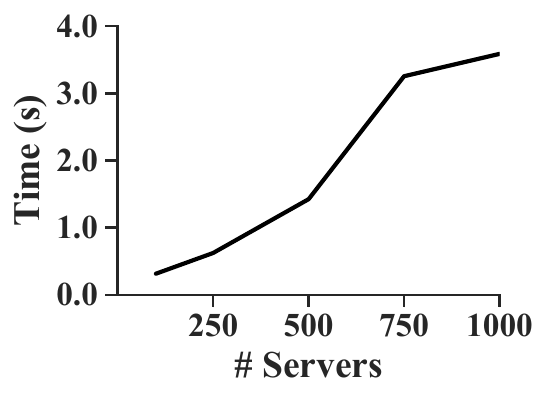}%
    \caption{\# Servers}%
    \label{fig:scalability_servers}%
  \end{subfigure}%
  \hfill%
  \begin{subfigure}[t]{0.15\textwidth}%
    \centering%
    \includegraphics[width=\textwidth]{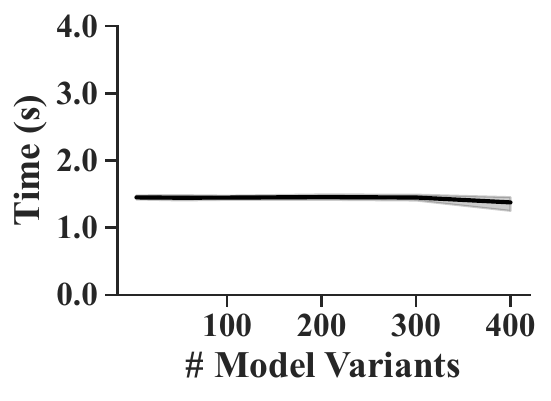}%
    \caption{\# Model Variants}%
    \label{fig:scalability_modelss}%
  \end{subfigure}%
  \caption{Scalability of \systemName heuristic model selection and placement. We fix the number of servers as 500, applications as 1000, and model variants as 4, and show the effect of changing each.
  }
  \label{fig:scalability}
  \vspace{-5mm}
\end{figure}

\subsection{\systemName Overheads}
In this section, we highlight the runtime overheads experienced by \systemName. Our failure detection process typically took 65ms per our configuration, while informing the clients with the location of the failover replica was around 10ms. 
\autoref{fig:scalability} highlights the scalability of our proposed heuristic, where even in very large scale settings (e.g., 3000 applications or 1000 servers), it took less than 4 seconds.  
Lastly, we note that the largest overhead source was the model loading time, which was not controlled by \systemName as it was a function of the model size as we shown in ~\autoref{fig:models_tradeoffs_loading}.  





%% file: sections/10-discussion.tex
\walid{We have shown the potential of \systemName in enhancing the resiliency of model serving systems in resource-constrained edge environments. In this section, we reflect on the generalizability and limitations of \systemName.
}

\noindent\textbf{Generalizability of \systemName.}
\walid{
While our experiments focus on enhancing the failure resilience of a single model serving system, \systemName's two-step failover process is generalizable across multiple scenarios. 
\systemName's can increase the resiliency of the DNN pipeline by prioritizing backups for critical applications, replicating different stages of DNN pipelines, or replicating a smaller version of such a pipeline.
Moreover, while \systemName is designed for resource-constrained edge environments, this paper lays out design principles that are beneficial in cloud environments where hosting failover replicas, although possible, is a costly option. In this case, employing heterogeneous replicas and progressively loading the failover model can bridge the gap between the high cost of proactive (always-on) failover replicas and the high MTTR of reactive failover replicas.  
}


\noindent\textbf{Limitations.}
\walid{While \systemName highly enhances the resiliency of model serving systems, \systemName depends on the availability of different model variants with different resource-accuracy trade-off points. Nonetheless, this assumption is justified as users typically train multiple sizes of the same model either intentionally~\cite{resnet, Tan2020:EfficientNet}, over time~\cite{Redmon2016:YOLO, Reis2024:YOLOv8}, using special architectures such as early exit architectures~\cite{Liang2023:Delen, yousefpour2019guardians, yousefpour2020resilinet, majeed2022continuer}, or using compression techniques, such as quantization and pruning \cite{Liu2020:Pruning, Hashemi2017:Quantization, Zadeh2020:GOBO}, all of which significantly decrease resource requirements and latency with a minor reduction in accuracy. Another limitation is that \systemName assumes that storage is abundant and can host various DNNs. However, we justify this assumption by the large differences between accelerator memory resources and storage, and the ability to use a cloud-based model repository to download models on the fly.
}





%% file: sections/8-related_work.tex
\noindent \textbf{Model Serving.} The importance of model serving systems has encouraged many researchers to address its latency ~\cite{Daniel2017:Clipper, Gujarati2020:Clockwork, Soifer2019:MSTInference, Zhang2019_HeteroEdge, Samplawski2020:Towards, Zhang2021:DeepSlicing, Liang2023:Queueing, Gujarati2020:Clockwork, Hanafy2023:Understanding}, cost \cite{Zhang202:MArk, Ahmad2024:Loki, Liang2020:AIEge, Fan2019:OnCost}, energy efficiency~\cite{Liang2023:Delen, Hanafy2021:DNNSelection, Wan2020:ALERT, Zadeh2020:GOBO}, and addressing challenges of workload dynamics~\cite{Ahmad2024:Proteus, Ahmad2024:Loki, Zhang2020:ModelSwitching, Wan2020:ALERT, Liang2023:Delen}. A common idea of these papers is exploring the accuracy-resource trade-offs where choosing the right-sized DNN model \cite{Hanafy2021:DNNSelection, Ahmad2024:Proteus, Zhang2020:ModelSwitching, Halpern2019:OneSize, Ahmad2024:Loki} or utilizing flexible DNN architecture (e.g., multi-exit DNNs ~\cite{Liang2023:Delen} and Slimmable DNNs~\cite{yu2018slimmable}). Although similar to many of these papers, \systemName exploits the accuracy-resource trade-offs, and we utilize it to enhance the resiliency of model serving systems and highlight a new accuracy-resiliency trade-off.

\noindent \textbf{Resilient Model Serving.}
\walid{Failure resilient model serving has been mostly addressed in the context of DNN training~\cite{He2023:DNNTrainingFail, Li2017:Understandingerror}.
However, little work has addressed failure resiliency in model serving systems~\cite{Soifer2019:MSTInference, yousefpour2019guardians, yousefpour2020resilinet, majeed2022continuer}. For instance, in \cite{Soifer2019:MSTInference}, the authors use active replication by replicating requests across servers to ensure resilient execution.   In contrast to this approach, \systemName addresses failure resiliency in resource-constrained environments, where such replication is not possible.
Another line of work~\cite{yousefpour2019guardians, yousefpour2020resilinet, majeed2022continuer} highlighted that as models become larger, the model inference systems will often utilize multiple nodes to process inference requests. In this case, the authors highlighted that special architectures, such as early-exit and skip-connection, can bring resilience by enabling early stages of the models to operate independently and produce results. Nonetheless, \systemName can utilize such architectures as they can be used as an alternative for model selection and employ failover replicas based on these architectures.  
}

\noindent \textbf{Resiliency in Resource Constraint Environments}
Graceful degradation is commonly used in scenarios with resource contention. For instance, Defcon~\cite{defcon} improves reliability by incorporating services' criticality and allocating resources per this criticality. In contrast to this paper, where placement decisions are binary, \systemName addresses the Failure Resiliency in model serving systems, where the decisions are far more complex and the system can utilize DNNs' flexibility by choosing different model sizes and selecting which applications have a failover replica.

%% file: sections/9-conclusion.tex
In this paper, we proposed \systemName, a failure-resilient model serving system that employs (i) a {\em heterogeneous replication}  where failover models are smaller variants of the original model, and  (ii) an intelligent approach that uses warm replicas to ensure quick failover for important applications while using cold replicas and (iii) {\em progressive failover} to provide low mean time to recovery for the remaining applications. 
Our evaluation has demonstrated that our two-step approach enables \systemName to maximize worst-case accuracy (i.e., accuracy during system failures) for critical and non-critical applications while minimizing MTTR. Our results using the 27 models show that FailLite can recover all failed applications with 175.5ms MTTR and only introduce 0.6\% reduction in accuracy. Our future work will address the failure resiliency in heterogeneous environments and when deploying model serving pipelines.